\pdfoutput=1
\documentclass{article}
\usepackage{spconf}
\usepackage{graphicx}
\usepackage{amsmath}
\usepackage{amssymb}
\usepackage{url}
\usepackage{tikz}
\usetikzlibrary{arrows.meta,positioning,fit,backgrounds}

\usepackage{booktabs}   
\usepackage{multirow}   
\usepackage{xcolor}
\usepackage{xspace}
\usepackage{array,colortbl,capt-of}
\usepackage{caption}
\usepackage{subcaption}
\DeclareRobustCommand{\systemname}{\textit{Lyric}\xspace}

\title{\textit{LYRIC}: Wave-Domain Computing for Efficient Spoken-Digit Recognition}

\name{Jeeven Balasubramaniam, Nakul Garg}
\address{Rice University}

\begin{document}
\ninept
\maketitle

\begin{abstract}
Low-power speech recognition requires reducing audio acquisition and feature-processing costs as well as neural inference.
We present \systemname{}, a speech-recognition front end that uses wave-domain computing to extract features before digitization.
Our proposed system uses passive acoustic resonators to separate speech by frequency and supplies their slowly varying envelopes to a compact temporal neural network.
This moves spectral filtering into the acoustic structure, reducing the data and processing needed for recognition.
We build a prototype and evaluate spoken-digit recognition on a speaker-disjoint AudioMNIST split across nine classifier families.
The front end acquires 32 times fewer scalar samples than a 16-kHz waveform.
In the EdgeSpeechNet-A comparison on a Raspberry Pi~4, we reduce preprocessing-plus-inference latency and estimated processor energy per inference by 98.6\%, with a 3.58-percentage-point decrease in accuracy to 95.70\%.
\end{abstract}


\section{Introduction}
\label{sec:intro}

Speech is becoming an everyday way to interact with computers, from giving commands to transcribing conversations.
Bringing these capabilities onto small, battery-powered devices remains energy intensive.
Conventional systems capture high-rate audio, compute spectral features, and run neural networks to recognize speech.
This pipeline preserves a detailed waveform before extracting the information needed to recover the words.
Yet a speech interface needs to understand what was said, without necessarily retaining everything needed to reproduce the sound.
In this paper, we ask: \emph{can we capture the information needed for speech recognition directly from sound, before high-rate digitization and digital feature extraction?}

\begin{figure}[t]
  \centering
  \includegraphics[width=\columnwidth]{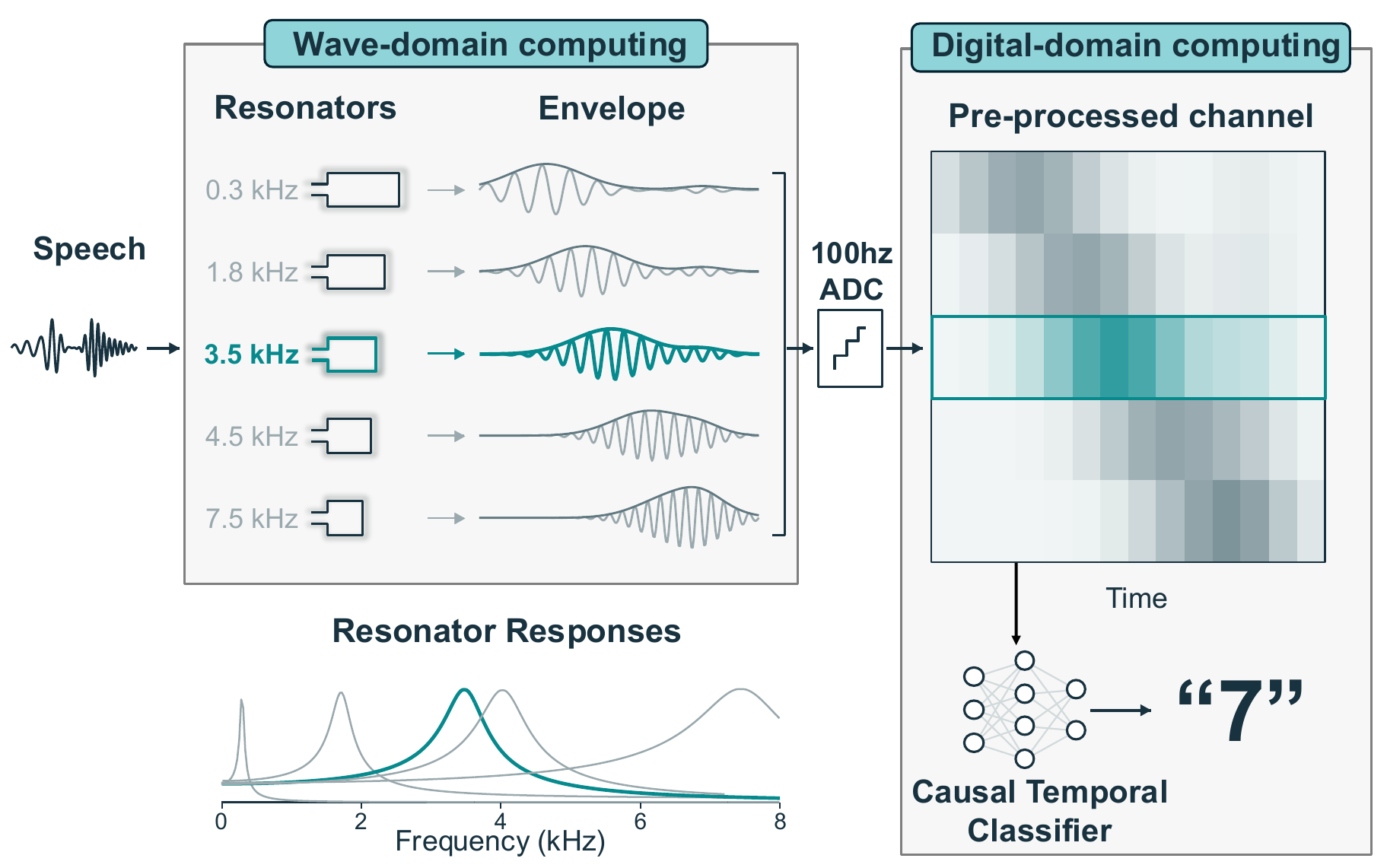}
  \vspace{-0.15in}
  \caption{\textbf{Speech recognition with wave-domain computing.}}
  \vspace{-0.2in}
  \label{fig:pipeline}
\end{figure}

Our proposed system, \systemname{}, uses \emph{wave-domain computing} to separate speech by frequency before digitization (Fig.~\ref{fig:pipeline}).
We build on the observation that, similar to the human cochlea, acoustic resonators respond selectively to different parts of the speech spectrum.
Their channel identities encode coarse frequency information, while their envelopes capture how the response amplitudes change over time.
We acquire these slower signals directly and train a temporal network to recognize the resulting patterns, without reconstructing the waveform or computing a digital spectrogram.
We use passive acoustics for frequency selection and a trainable temporal network for recognition.

We design the front end around a small set of channels that retains useful differences between speech sounds.
Using physics-based simulations and a separate set of alphabet recordings, we compare candidate channel combinations before evaluating on the target digit dataset.
We build the selected five channels with 3D-printed Helmholtz resonators, microphones and analog envelope readout.
At 100 samples/s per channel, the prototype produces 32 times fewer scalar samples than a 16-kHz waveform while preserving the 100-Hz update rate of the log-Mel baseline.

We build on prior demonstrations of physical frequency separation and speech recognition~\cite{dalpoggetto2023tonotopic,dubcek2024phononic,beoletto2026speech}.
Related work has also explored analog filter-bank feature extraction for ultra-low-power keyword spotting~\cite{ray2023tiny,shen2025kws}.
Our focus is the compact representation passed from acoustics to a trainable recognizer and the processing cost it removes.
This complements pruning and quantization~\cite{han2015compression,jacob2018quantization}, which reduce the cost of the network itself.

We evaluate \systemname{} on physically recorded AudioMNIST utterances using a speaker-disjoint split and nine classifier families~\cite{becker2024audiomnist}.
The measured responses show that digits differ both in which channels respond and in when those responses occur.
In our ablations, a single broadband envelope gives substantially lower accuracy, while increasing the five-channel envelope rate from 100 to 400~Hz provides no consistent accuracy gain across the tested models.
These observations support retaining frequency-selective temporal patterns while reducing how much audio the processor receives.

We measure the resulting accuracy--computation tradeoff against log-Mel inputs using representation-specific models from the same classifier families.
The seven larger envelope models achieve 91.07--95.70\% digit accuracy, and all nine families reduce preprocessing-plus-inference latency on a Raspberry Pi~4.
For EdgeSpeechNet-A, accuracy changes from 99.28\% to 95.70\%, while computation latency and estimated processor energy per inference fall by 98.6\% under the same processor-power model.
We thus demonstrate a 3.58-percentage-point accuracy tradeoff for substantially lower digital processing cost in this spoken-digit test.


\section{Problem Formulation and Intuitions}
\label{sec:problem}

Our problem is to recognize speech from sparse measurements produced by the acoustic front end.
To solve this, we seek a latent representation that compresses the analog signals while preserving features between digits spoken by previously unseen speakers.

\noindent\textbf{Intuition.}
We draw inspiration from the human cochlea's separation of sound into frequency-dependent responses~\cite{lyon1988analogcochlea}.
Human listening experiments also show that speech can remain intelligible when temporal envelopes are retained in a few broad frequency bands~\cite{shannon1995temporal}.
In our resonators, air moving in the neck compresses the air in the cavity while their geometry determines which frequencies produce a strong response.
The envelopes then describe how energy in these frequency regions changes through the sound.
Our hypothesis is that we can engineer these resonance responses to retain enough information for recognition even when the detailed waveform is discarded.

Physical frequency encoding already has close precedents, so we focus on the representation and resource tradeoff it enables.
Prior work has demonstrated recurrent computation through simulated wave propagation and a tunable nonlinear acoustic cavity~\cite{hughes2019wavephysics,momeni2023neuroacoustic}, as well as passive phononic structures designed for pairwise word discrimination~\cite{dubcek2024phononic}.
Cochlea-inspired approaches include a 3D-printed tonotopic spiral~\cite{dalpoggetto2023tonotopic}, an artificial basilar-membrane system for keyword recognition~\cite{lee2025artificial}, and an event-driven silicon-cochlea front end for low-compute keyword spotting~\cite{ceolini2019event}.
Beoletto et al. extended mechanically frequency-selective sensing to digit recognition using a piezoelectrically driven spiral, 16 readout channels, and fixed-size tonograms~\cite{beoletto2026speech}.
More broadly, physical computing has recently been demonstrated for analogue speech recognition using in-materia temporal-signal processing~\cite{zolfagharinejad2025analogue}.
We use five microphone-coupled resonators driven directly by airborne speech, read their envelopes at low rate, and evaluate both recognition accuracy and digital processing cost.
Our contribution is the implementation and evaluation of this compact acoustic-to-digital representation, rather than the principle of physical frequency separation alone.

\noindent\textbf{Challenges.}
We must choose channels that preserve complementary speech cues without sacrificing their timing.
Channels that respond similarly provide redundant information, while gaps in coverage can remove distinctions between digits.
A sharply tuned resonator can continue ringing after the sound changes, and envelope smoothing can blur onsets and transitions.
We therefore select channels using their simulated speech responses and retain a temporal network to combine the measured envelopes over the utterance.
The envelope signals still require appropriate filtering and sampling; only acoustic frequency selection is passive.
We next describe the physics-based simulations, resonator realization, and temporal recognition pipeline used to test this design.



\section{System Design}
\label{sec:method}

\subsection{Design Overview}
\label{ssec:overview}

\systemname{} uses acoustic filtering and analog envelope extraction to replace digital log-Mel computation (Fig.~\ref{fig:pipeline}).
Airborne speech drives five passive Helmholtz resonators, each tuned to a different region of the speech spectrum.
A microphone sealed into each cavity measures its pressure response.
An analog AC-RMS circuit extracts the slowly varying amplitude over a 10-ms integration interval.
The five envelopes are digitized at 100 samples/s per channel and passed to a causal temporal network.
This path requires no digital framing, FFT, Mel filtering, or logarithmic compression.

We first select five channels using simulated responses to alphabet speech (Sec.~\ref{ssec:comsol_selection}), then fabricate and adjust the resonators (Sec.~\ref{ssec:resonator_realization}).
We examine the spectral and temporal structure retained by the array in Sec.~\ref{ssec:representation} and account for its acquisition and processing costs in Sec.~\ref{ssec:cost}.

\subsection{Choosing the Channels}
\label{ssec:comsol_selection}

Channel selection must balance redundancy against spectral coverage.
Channels with similar responses can carry overlapping information, while gaps in coverage can discard differences between sounds.
We therefore screened candidate banks in simulation using a separate alphabet dataset, before recording any AudioMNIST data.

We simulated eight candidate resonators in COMSOL at target frequencies of 110, 180, 300, 1800, 3500, 4500, 6000, and 7500\,Hz.
Let $v_i(f)$ denote the complex sensor velocity of candidate $i$ and $p_{\mathrm{in}}(f)$ the complex inlet pressure exported by the frequency-domain simulation.
We define the peak-normalized transfer-energy response as
\begin{equation}
  G_i(f)=
  \frac{\left|v_i(f)/p_{\mathrm{in}}(f)\right|^2}
       {\max_{f'}\left|v_i(f')/p_{\mathrm{in}}(f')\right|^2}.
  \label{eq:comsol_transfer}
\end{equation}
For an utterance with short-time power spectrum $S_x(f,t)$, the predicted response trajectory is $r_i(t)=\sum_f G_i(f)\,S_x(f,t)$, summed over 80\,Hz--8\,kHz.
We compute $S_x(f,t)$ from 16-kHz speech using a 40-ms Hann window, a 10-ms hop, and a 2048-point FFT.
These 100-Hz trajectories provide a spectral-energy proxy for channel selection.

We evaluated all $\binom{8}{5}=56$ five-resonator subsets on 182 labeled A--Z segments from seven recordings.
Each subset was represented by 25 features describing per-channel energy proportions and their temporal changes.
We used a class-balanced linear SVM ($C=0.3$) with seven-fold leave-one-recording-out cross-validation, fitting standardization only on each fold's training recordings.

The subset targeting 300, 1800, 3500, 4500, and 7500\,Hz ranked first at 67.03\% mean balanced accuracy, compared with 66.48\% for the runner-up.
Its fold standard deviation was 15.23 percentage points.
We selected the highest-scoring bank, but the narrow margin and large fold variation do not establish a unique optimum.
The selection used alphabet speech only, so no digit data informed the channel choice.
We denote a channel targeting frequency $f$ by R$f$, giving R300, R1800, R3500, R4500, and R7500.
\vspace{-0.1in}

\subsection{Building the Resonators}
\label{ssec:resonator_realization}

\begin{figure}[!htbp]
  \centering
  \vspace{-0.1in}
  \includegraphics[width=0.8\columnwidth]{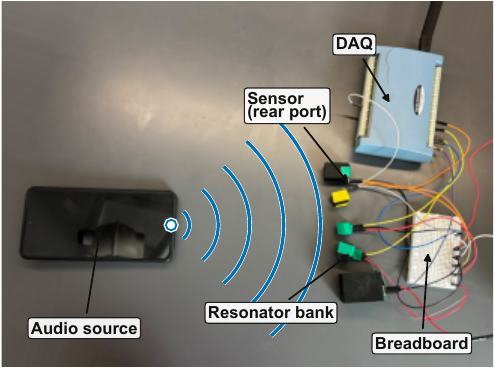}
  \vspace{-0.1in}
  \caption{Our setup for fabricated five-channel \systemname{} resonator bank.}
  \label{fig:real_resonators}
  \vspace{-0.1in}
\end{figure}

Each channel is a monolithic PLA Helmholtz resonator (Fig.~\ref{fig:real_resonators}).
We print the neck, cavity, transition, and microphone port as one part to eliminate assembly seams that could alter the cavity response.

For a neck of area $A$ and effective length $L_{\mathrm{eff}}$ coupled to a total acoustic volume $V$, the first-order design relation is
\begin{equation}
  f_{\mathrm{H}}=\frac{c}{2\pi}\sqrt{\frac{A}{V L_{\mathrm{eff}}}},
  \label{eq:helmholtz}
\end{equation}
where $c$ is the speed of sound and $L_{\mathrm{eff}}=L+0.85\,d$ accounts for the end correction of a neck with physical length $L$ and diameter $d$.
Table~\ref{tab:resonators} lists the final geometries, with total acoustic volumes ranging from 0.077 to 9.213\,mL.

\begin{table}[!htbp]
\centering
\fontsize{9}{10}\selectfont
\setlength{\tabcolsep}{3.5pt}
\begin{tabular}{rrrrrrr}
\toprule
$f_{\mathrm{target}}$ & \multicolumn{3}{c}{Neck (mm)} & \multicolumn{2}{c}{Cavity (mm)} & $V$ \\
\cmidrule(lr){2-4}\cmidrule(lr){5-6}
(Hz) & $d$ & $L$ & $L_{\mathrm{eff}}$ & Side & Depth & (mL) \\
\midrule
300  & 2.50 & 16.28 & 18.41 & 23.46 & 12.00 & 9.213 \\
1800 & 3.39 &  4.00 &  6.88 &  8.87 & 15.36 & 1.404 \\
3500 & 6.00 & 10.00 & 15.10 & 12.16 &  3.20 & 0.475 \\
4500 & 5.00 &  6.40 & 10.65 &  8.01 &  3.04 & 0.283 \\
7500 & 4.00 &  4.00 &  7.40 &  6.84 &  1.60 & 0.077 \\
\bottomrule
\end{tabular}
\vspace{-0.1in}
\caption{The final resonator geometries.}
\label{tab:resonators}
\vspace{-0.2in}
\end{table}

Equation~(\ref{eq:helmholtz}) provides an initial geometry.
Fabrication tolerances, acoustic loading, and differences in effective neck length can shift the realized resonance.
Where prototype measurements were available, we used the measured peak to adjust the cavity volume while holding the neck fixed:
\begin{equation}
  V_{\mathrm{new}}=V_{\mathrm{old}}
  \left(\frac{f_{\mathrm{meas}}}{f_{\mathrm{target}}}\right)^{2}.
  \label{eq:volume_correction}
\end{equation}
For the 1800-Hz channel, a predecessor peaked at 2150\,Hz.
Applying this correction increased its volume from 0.984 to 1.404\,mL.

\subsection{What the Array Measures}
\label{ssec:representation}

\begin{figure}[htbp]
  \centering
  \vspace{-0.1in}
  \includegraphics[width=0.85\columnwidth]{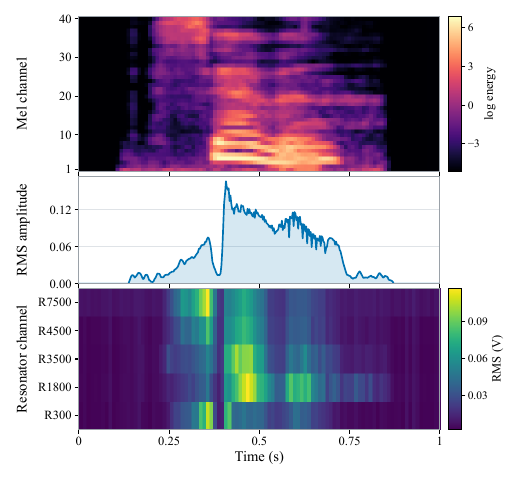}
  \vspace{-0.2in}
  \caption{Representations of one spoken ``seven'': log-Mel, broadband envelope, and \systemname{}, from top to bottom.}
  \label{fig:frontend_comparison}
  \vspace{-0.1in}
\end{figure}

Fig.~\ref{fig:frontend_comparison} compares the measured five-channel stream with the digital representations evaluated in Sec.~\ref{sec:results}.
The broadband envelope retains amplitude changes over time but does not separate frequency bands.
The log-Mel representation retains finer spectral structure.
The resonator array retains a coarser spectral representation, separating five regions and tracking their amplitudes over time.
Sec.~\ref{ssec:ablations} will test the effects of retaining separate channels and varying sampling rates.

\begin{figure}[t]
  \centering
  \vspace{-0.1in}
  \includegraphics[width=\columnwidth]{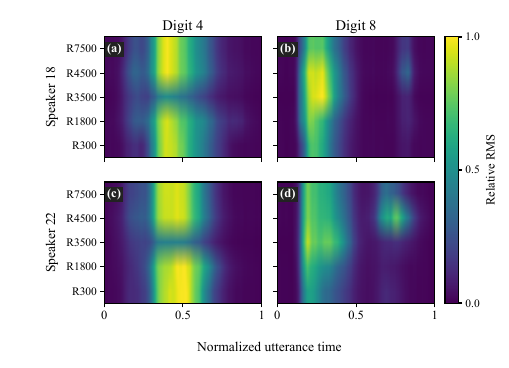}
  \vspace{-0.3in}
  \caption{Measured five-channel envelope templates for ~digit~4 and ~8 from different speakers, averaged over 50 repetitions per class.}
  \label{fig:templates}
  \vspace{-0.25in}
\end{figure}

Fig.~\ref{fig:templates} shows class templates averaged over 50 repetitions through the fabricated array.
For ``four,'' the response contains a sustained mid-utterance burst, with strong R300, R1800, R4500, and R7500 responses and a weaker R3500 response.
For ``eight,'' an early burst is concentrated in R3500--R4500 in both speakers; a separate late high-frequency burst is more visible for speaker~22.
The templates differ in both channel activity and event timing.
We retain a temporal network to use this timing structure, which would be lost by pooling each channel over the utterance.

Resonator ringing and envelope integration can blur the onsets and transitions that distinguish short speech sounds.
A resonator continues to respond after excitation ends, and the 10-ms RMS integration further smooths its output.
The envelope integration interval is shorter than the baseline's 50-ms log-Mel window, with both representations using a 10-ms hop.
This difference alone does not establish the array's temporal resolution because we did not measure the ring-down of each fabricated channel.
The separated bursts visible in Fig.~\ref{fig:templates}, particularly for speaker~22's ``eight,'' show that these events remain distinguishable in the averaged envelopes; they do not provide a measured bound on transient response.
\vspace{-0.1in}

\subsection{Acquisition and Processing Cost}
\label{ssec:cost}
\label{ssec:frontend}

The conventional front end used for comparison~\cite{sainath2015,choi2019,kim2021} acquires 16,000 waveform samples/s and computes a 40-bin log-Mel spectrogram using a 50-ms window and a 10-ms hop.
This produces 100 frames/s.
Before classification, each overlapping frame requires windowing, an FFT, power-spectrum computation, Mel filtering, logarithmic compression, and normalization.
Reducing the classifier size does not remove this preprocessing.

\systemname{} performs frequency selection and extracts the envelopes before digitization.
The ADCs therefore sample five amplitude signals at 100 samples/s each: 500 scalar samples/s in total.
This is $32\times$ fewer samples than the 16-kHz waveform and one-eighth the number of log-Mel feature values, at the same 100-Hz frame rate.
Channel identity carries the coarse spectral position, so the ADCs do not need to sample the acoustic carriers.
The 50-Hz Nyquist limit applies to \emph{envelope modulation}, while the channel target frequencies extend to 7.5\,kHz.
Modulation above 50\,Hz must still be attenuated by the analog envelope response before digitization.

\section{Evaluation}
\label{sec:results}

\subsection{Experimental Setup}
\label{ssec:setup}
\label{ssec:data_collection}
\label{ssec:classifiers}

We evaluate on AudioMNIST~\cite{becker2024audiomnist}, with 30,000 digit recordings split into 36 training, 12 validation, and 12 test speakers.
Audio was peak-normalized to $-9$\,dBFS and replayed 15\,cm from the fixed resonator array, one speaker per session, with 150-ms gaps.
The five channels were recorded simultaneously and segmented using the playback schedule.
Normalization uses training statistics only; test replay followed model selection.

We compare 100-Hz resonator envelopes with 40-bin log-Mel features from 16-kHz audio (Sec.~\ref{ssec:frontend}).
Digital recordings are normalized and cropped or padded to 1\,s.
For ablations we also use a 400-Hz resonator variant with a 10-ms window and a 2.5-ms hop, and a single broadband RMS envelope computed from the unfiltered waveform with the same window and rate but no resonator filtering.
The classifiers are MicroCNN~\cite{tflmMicroSpeech}, StrideCNN and PoolCNN~\cite{sainath2015}, DS-CNN~\cite{banbury2021}, TC-ResNet8~\cite{choi2019}, BC-ResNet-1/2~\cite{kim2021}, MatchboxNet-3$\times$1$\times$64~\cite{majumdar2020}, and EdgeSpeechNet-A~\cite{lin2018}.
Envelope models use causal 1-D adaptations of these families and length-masked mean logits:
\begin{equation}
\hat{\mathbf{y}}=\frac{1}{T'}\sum_{t=1}^{T'}\mathbf{z}_t,\qquad T'=\lceil T/s\rceil,
\label{eq:masked_mean}
\end{equation}
where $T$ counts valid input frames and $s$ is temporal stride.
Paired models retain the classifier family's main structure but differ in input representation and architecture.
The comparison therefore measures the complete pipeline tradeoff rather than isolating the effect of physical filtering.

Training uses cross-entropy and AdamW with learning rate $10^{-3}$ and weight decay $10^{-4}$.
Each configuration uses one seed; validation accuracy selects the checkpoint, with validation loss breaking ties.
Resonator/broadband models use batch sizes 128/256 and up to 180/50 epochs.
We report median preprocessing-plus-inference time over 6,000 utterances on a Raspberry Pi~4 (1.8\,GHz, 8\,GB, performance governor, one thread, batch size~1, PyTorch~2.14).
This measures processor computation, excluding acquisition and the wait for the complete utterance.

\vspace{-0.1in}

\begin{table}[t]
\centering
\captionsetup{font=small,skip=3pt,labelfont=bf,labelsep=period}
\definecolor{evalgood}{HTML}{FBE5D3}
\fontsize{8.5}{10}\selectfont
\setlength{\tabcolsep}{2.5pt}
\renewcommand{\arraystretch}{1.20}
\newcommand{\good}[1]{\cellcolor{evalgood}#1}

\begin{tabular}{@{}l rrr rrr@{}}
\toprule
& \multicolumn{3}{c}{log-Mel}
& \multicolumn{3}{c}{\systemname{}} \\
\cmidrule(lr){2-4}\cmidrule(l){5-7}
Model & Params & Acc. & Time & Params & Acc. & Time \\
& (k) & (\%) & (ms) & (k) & (\%) & (ms) \\
\midrule
MicroCNN~\cite{tflmMicroSpeech}
& 0.7 & 43.88 & 13.41 & 0.5 & 74.72 & 1.29 \\
StrideCNN~\cite{sainath2015}
& 1.2 & 47.52 & 13.59 & 0.8 & 80.92 & 1.39 \\
BC-ResNet-1~\cite{kim2021}
& 9.2 & 98.42 & 91.88 & 9.1 & 92.07 & 15.59 \\
PoolCNN~\cite{sainath2015}
& 13.0 & 96.00 & 32.22 & 4.4 & \good{91.07} & \textbf{\good{1.81}} \\
DS-CNN~\cite{banbury2021}
& 23.1 & 99.02 & 55.31 & 22.2 & \good{94.47} & \textbf{\good{6.21}} \\
TC-ResNet8~\cite{choi2019}
& 25.5 & 98.60 & 16.87 & 23.9 & \good{93.57} & \textbf{\good{3.83}} \\
BC-ResNet-2~\cite{kim2021}
& 27.2 & 98.55 & 109.50 & 31.8 & 92.78 & 16.26 \\
MatchboxNet~\cite{majumdar2020}
& 62.9 & 99.10 & 30.37 & 58.0 & \good{93.93} & \textbf{\good{9.26}} \\
EdgeSpeechNet-A~\cite{lin2018}
& 108.1 & 99.28 & 454.31 & 37.4 & \good{95.70} & \textbf{\good{6.56}} \\
\bottomrule
\end{tabular}

\caption{AudioMNIST accuracy and compute cost across nine classifier families using log-Mel and \systemname{} inputs.
Highlighted configurations achieve $\geq90\%$ accuracy in $<10$\,ms.}
\vspace{-0.2in}
\label{tab:results}
\end{table}

\subsection{Recognition Accuracy and Processing Cost}
\label{ssec:accuracy}
\label{ssec:accuracy_cost}
\label{ssec:latency_energy}
\label{ssec:pareto}

\systemname{} achieves 95.70\% accuracy with EdgeSpeechNet-A at 6.56\,ms and an estimated 1.97\,mJ of dynamic processor energy per inference, compared with 99.28\%, 454.31\,ms, and 136.29\,mJ for its log-Mel counterpart (Table~\ref{tab:results}).
Compute time and estimated energy fall by approximately $69\times$, with a 3.58-percentage-point accuracy loss.
Across the seven families exceeding 95\% with log-Mel, the five-channel input achieves 91.07--95.70\% accuracy.
The two smallest classifiers improve instead: MicroCNN rises from 43.88\% to 74.72\%, and StrideCNN from 47.52\% to 80.92\%, despite fewer parameters.

Replacing digital log-Mel extraction with envelope normalization reduces mean front-end time from 8.48 to 0.28\,ms.
The five-channel input and its adapted networks also require fewer MACs in every pair, while retaining the 100-Hz input frame rate.
Together, these changes reduce total compute latency by 69.5--98.6\% across the nine families.

\begin{figure}[t]
\centering
\captionsetup{font=small,skip=3pt,labelfont=bf,labelsep=period}
\vspace{-0.15in}
\includegraphics[width=0.9\columnwidth,trim=0 8bp 0 0,clip]{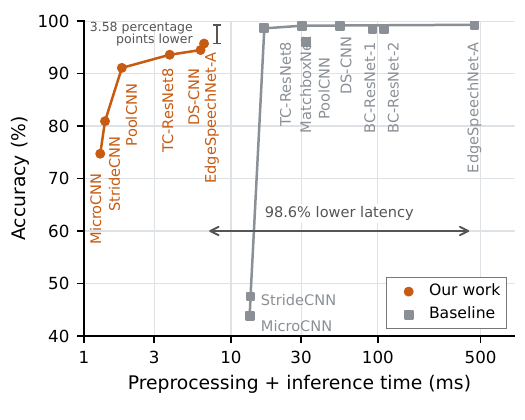}
\caption{Recognition accuracy versus latency for log-Mel and \systemname{} inputs.
Arrows mark the paired EdgeSpeechNet-A comparison.}
\vspace{-0.15in}
\label{fig:pareto}
\end{figure}

Five \systemname{} configurations exceed 90\% accuracy in under 10\,ms and no evaluated log-Mel configuration meets both thresholds (shaded cells in Table~\ref{tab:results}).
Resonator PoolCNN reaches 91.07\% at 1.81\,ms and 0.54\,mJ; the cheapest log-Mel model above 90\%, TC-ResNet8, requires 16.87\,ms and 5.06\,mJ (Fig.~\ref{fig:pareto}).
For higher accuracy, resonator EdgeSpeechNet-A is 2.9 points below log-Mel TC-ResNet8, with a $2.6\times$ reduction in compute time and estimated energy.
The gain thus remains against a lower-cost log-Mel baseline, although it is smaller than the paired EdgeSpeechNet-A reduction.
For fair comparison of computation cost, energy estimates use the measured 0.30-W dynamic processor draw and exclude microphones, envelope circuitry, ADC, and acquisition time.

\subsection{Ablations}
\label{ssec:ablations}

\noindent\emph{Envelope rate.}
\label{ssec:rate_ablation}
Increasing the resonator rate from 100 to 400\,Hz raises MACs $3.8$--$4.0\times$ and total latency $2.3$--$4.6\times$, but changes mean accuracy across the seven stronger families only from 93.37\% to 93.59\%.
MicroCNN and StrideCNN instead lose 7.45 and 4.57 points.
The 100-Hz input avoids this additional compute cost without a consistent loss of accuracy in these runs.
Single-seed results cannot quantify run-to-run variation.

\noindent\emph{Spectral channels.}
\label{ssec:ablation}
At 400\,Hz, using identical 10-ms windows and 2.5-ms hops, five resonator channels outperform a single broadband RMS envelope by 21.1--44.9 percentage points across the tested architectures.
Different training budgets limit attribution of this gain to spectral selectivity.
This comparison does not establish an advantage over an equivalent five-channel digital filter bank.

\section{Conclusion}
\label{sec:conclusion}

We presented \systemname{}, which uses wave-domain computing to extract speech features before digitization.
Our spoken-digit experiments show that low-rate resonator envelopes preserve useful spectral and temporal cues, enabling recognition with substantially less digital processing than a log-Mel pipeline.
Future work will integrate the sensing and inference hardware, measure total system energy, and evaluate the approach on continuous speech.


\bibliographystyle{IEEEtran}
\bibliography{refs}

\end{document}